%% file: hadron2011.tex
\documentclass[a4paper,11pt]{article}

\usepackage{contribution}
\usepackage{epstopdf}
\input{econfmacros}

\input{contributionmacros}

\begin{document}

% % % % % % % % % % % % % % % % % % % % % % % % % % % % % % % % % % % % % % % % %
% your proceedings
\input{contribution}

\end{document}

%% file: econfmacros.tex
\newcommand{\weblink}[2][]{%
    \ifthenelse{\equal{#1}{}}%
    {\textnormal{\url{#2}}}%
    {\textnormal{\href{#2}{#1}}}%
}

\newcommand{\acknowledgements}[1]{%
  \bigskip\bigskip
  \textsf{\textbf{\Large Acknowledgements}} \\[2ex]
  {#1}
  \bigskip
}

\def\beq{\begin{equation}}
\def\eeq#1{\label{#1}\end{equation}}
\def\eeqn{\end{equation}}

\def\beqa{\begin{eqnarray}}
\def\eeqa#1{\label{#1}\end{eqnarray}}
\def\eeqan{\end{eqnarray}}

\let\bar=\overbar

\def\etal{{\it et al.}}

\def\Dslash{\not{\hbox{\kern-4pt $D$}}}
\def\dslash{\not{\hbox{\kern-2pt $\del$}}}

\def\msb{{\bar{\ssstyle M \kern -1pt S}}}

%% file: contributionmacros.tex
\newcommand{\contribution}[7][]{%
  \clearpage
  \thispagestyle{plain}
  \ifthenelse{\equal{#1}{}}
  {\hypersetup{pdftitle={#2}}}
  {\hypersetup{pdftitle={#1}}}
  \hypersetup{pdfauthor={{#3} {#4}}}
  {\centering\normalfont\LARGE\bfseries\sffamily #2 \par\nobreak}
  \lhead{}
  \chead{%
    \textit{\footnotesize XIV International Conference on Hadron Spectroscopy
      (\weblink[\textit{hadron2011}]{http://www.hadron2011.de}), 13-17 June 2011, Munich, Germany}%
  }
  \rhead{}
  \bigskip
  \begin{center}
    {#3} {#4}\ifthenelse{\equal{#6}{}}{}{\footnote{\weblink[#6]{mailto:#6}}}
    \ifthenelse{\equal{#7}{}}{}{#7} \\
    \textit{#5}
  \end{center}
  \bigskip
}

\renewcommand{\abstract}[1]{%
  \begin{center}
    \begin{minipage}{0.85\textwidth}
      \begin{footnotesize}
        #1
      \end{footnotesize}
    \end{minipage}
  \end{center}
  \bigskip
}

%% file: contribution.tex
%%%%%%%%%%%%%%%%%%%%%%%%%%%%%%%%%%%%%%%%%%%%%%%%%%%%%%%%%%%%%%%%%%%%%%%%%%%%%%%%%
%
% template for hadron2011 contribution
%
% please do not rename this file
%
% to create document run
%
%     pdflatex hadron2011.tex
%
%%%%%%%%%%%%%%%%%%%%%%%%%%%%%%%%%%%%%%%%%%%%%%%%%%%%%%%%%%%%%%%%%%%%%%%%%%%%%%%%%
{  % do not remove

%%%%%%%%%%%%%%%%%%%%%%%%%%%%%%%%%%%%%%%%%%%%%%%%%%%%%%%%%%%%%%%%%%%%%%%%%%%%%%%%%
% please define your macros here
\makeatletter
\@ifundefined{c@affiliation}%
{\newcounter{affiliation}}{}%
\makeatother
\newcommand{\affiliation}[2][]{\setcounter{affiliation}{#2}%
  \ensuremath{{^{\alph{affiliation}}}\text{#1}}}

%
%%%%%%%%%%%%%%%%%%%%%%%%%%%%%%%%%%%%%%%%%%%%%%%%%%%%%%%%%%%%%%%%%%%%%%%%%%%%%%%%%

%%%%%%%%%%%%%%%%%%%%%%%%%%%%%%%%%%%%%%%%%%%%%%%%%%%%%%%%%%%%%%%%%%%%%%%%%%%%%%%%%
% define title, author, and address
% contribution[short title]{title}{author first name}{author last name}{author address}{author email}{collaboration}
% the short title will appear in the page headers and the TOC of the book of proceedings
% the last two arguments may be left empty
\contribution[The light nuclei spin structure]  % short title (optional)
{The light nuclei spin structure from hadronic \\ channels at intermediate energies}  % title
{P. K.}{Kurilkin}  % presenter of the talk/poster
{\affiliation[Joint Institute for Nuclear Research, Dubna, Russia]{1} \\
 \affiliation[Center for Nuclear Study, University of Tokyo, Tokyo, Japan]{2} \\
 \affiliation[Department of Physics, Saitama University, Urawa, Japan]{3}\\
 \affiliation[Physics Department, University of Zilina, Zilina, Slovakia]{4}\\
 \affiliation[Advanced Research Institute for Electrical Engineering, Bucharest, Romania]{5}\\
 \affiliation[Institute for Nuclear Research, Moscow, Russia]{6}\\
 \affiliation[Kyushi University, Harozaki, Japan]{7}\\
 \affiliation[P.J.Safarik University, Kosice, Slovakia]{8}\\
 \affiliation[University of Chemical Technology and Metallurgy, Sofia, Bulgaria]{9}\\
 \affiliation[RIKEN (the Institute for Physical and Chemical Research), Saitama, Japan]{10}\\
 \affiliation[University of Tokyo, Tokyo, Japan]{11}\\
 \affiliation[Belgorod State University, Belgorod, Russia]{12}
}
 {pkurilkin@jinr.ru}
{\!\!$^,\affiliation{1}$, V.P. Ladygin\affiliation{1}, T. Uesaka\affiliation{2}, V.V. Glagolev\affiliation{1}, Yu.V. Gurchin\affiliation{1}, A.Yu. Isupov\affiliation{1}, K. Itoh\affiliation{3}, M. Janek\affiliation{1}$^,$\affiliation{4}, J.-T. Karachuk\affiliation{1}$^,$\affiliation{5}, T. Kawabata\affiliation{2}, A.N. Khrenov\affiliation{1}, A.S. Kiselev\affiliation{1}, V.A.~Kizka\affiliation{1}, A.B. Kurepin\affiliation{6}, A.K. Kurilkin\affiliation{1}, V.A. Krasnov\affiliation{1}$^,$\affiliation{6}, N.B. Ladygina\affiliation{1}, D. Lipchinski\affiliation{5}, A.N. Livanov\affiliation{1}$^,$\affiliation{6}, Y. Maeda\affiliation{7}, A.I. Malakhov\affiliation{1}, G. Martinska\affiliation{8}, S. Nedev\affiliation{9}, S.M. Piyadin\affiliation{1},  E.B.~Plekhanov\affiliation{1}, J. Popovichi\affiliation{5}, S. Rangelov\affiliation{9}, S.G. Reznikov\affiliation{1}, P.A. Rukoyatkin\affiliation{1}, S.~Sakaguchi\affiliation{10}, H.~Sakai\affiliation{3}$^,$\affiliation{11}, K. Sekiguchi\affiliation{10}, K. Suda\affiliation{10}, A.A. Terekhin\affiliation{1}$^,$\affiliation{12}, J. Urban\affiliation{8}, T.A. Vasiliev\affiliation{1} and I.E. Vnukov\affiliation{12}
}

%%%%%%%%%%%%%%%%%%%%%%%%%%%%%%%%%%%%%%%%%%%%%%%%%%%%%%%%%%%%%%%%%%%%%%%%%%%%%%%%%
% abstract
\abstract{%
   The investigation of the d, $^3$H and $^3$He spin structure has been performed at the RIKEN(Japan) accelerator research facility and VBLHEP(JINR) using both polarized and unpolarized deuteron beams. The experimental results on the analyzing powers studies in $dp-$ elastic scattering, d(d,$^3$H)p and d(d,$^3$He)n reactions are presented. The vector and tensor analyzing powers for $dp-$ elastic scattering at 880 and 2000 MeV are obtained at the Nuclotron(VBLHEP). The result on the analyzing powers $A_{y}$, $A_{yy}$ of the deuteron  at 2000 MeV are compared with relativistic multiple scattering model calculations. The data on the tensor analyzing powers for the d(d,$^3$H)p and d(d,$^3$He)n reactions obtained at $E_{d}=200$ and 270 MeV demonstrate the sensitivity to the $^3$H, $^3$He and deuteron spin structure. The essential disagreements between the experimental results and the theoretical calculations within the one-nucleon exchange model framework are observed. 
The wide experimental program on the study of the polarization effects in $dp-$ elastic scattering, $dp-$nonmesonic breakup, d(d,$^3$He)n, d(d,$^3$H)p and d($^3$He,$^4$He)p reactions using internal and extracted beam at Nuclotron is discussed.
}
%
% main text
% for short contributions sections are optional
\section{Introduction}
The main goal of the investigation of the reaction induced by the polarized deuterons is to establish the nature of $2N$ and $3N$ forces, the role of the relativistic effects and nonucleon degrees of freedom. The last decades such investigation were performed at different experiments all over the world at RIKEN, KVI, IUCF and RCNP.
This activity was stimulated by the discrepancy of $30\%$ between the measured cross section for deuteron-proton($dp-$) elastic scattering at intermediate energies and the Faddeev calculations using modern potentials of nucleon-nucleon interaction. 
\section{Experimental results}
The research program on the light nuclei structure investigation at the Nuclotron includes experiments using both internal and extracted polarized deuteron beams.\\
The study of the energy dependence of polarization observables for the $dp-$ elastic scattering and deuteron breakup reaction are conducted at internal target station(ITS) setup. A detailed description of the experiment can be found in \cite{its_experiment}.\\ 
The deuteron analyzing powers measurements in $dp$- elastic scattering have been performed at ITS using polarized beam from polarized ion source (PIS) POLARIS at the energies 880 and 2000 MeV. The beam polarization measurement has been performed at 270 MeV where the precise data on the tensor and vector analyzing powers exist.\\
\begin{figure}[b]
\begin{minipage}[t]{0.48\textwidth}
  \centering
    % please do not add file name extension this makes switching between latex and pdflatex easier
    \resizebox{0.97\textwidth}{!}{\includegraphics{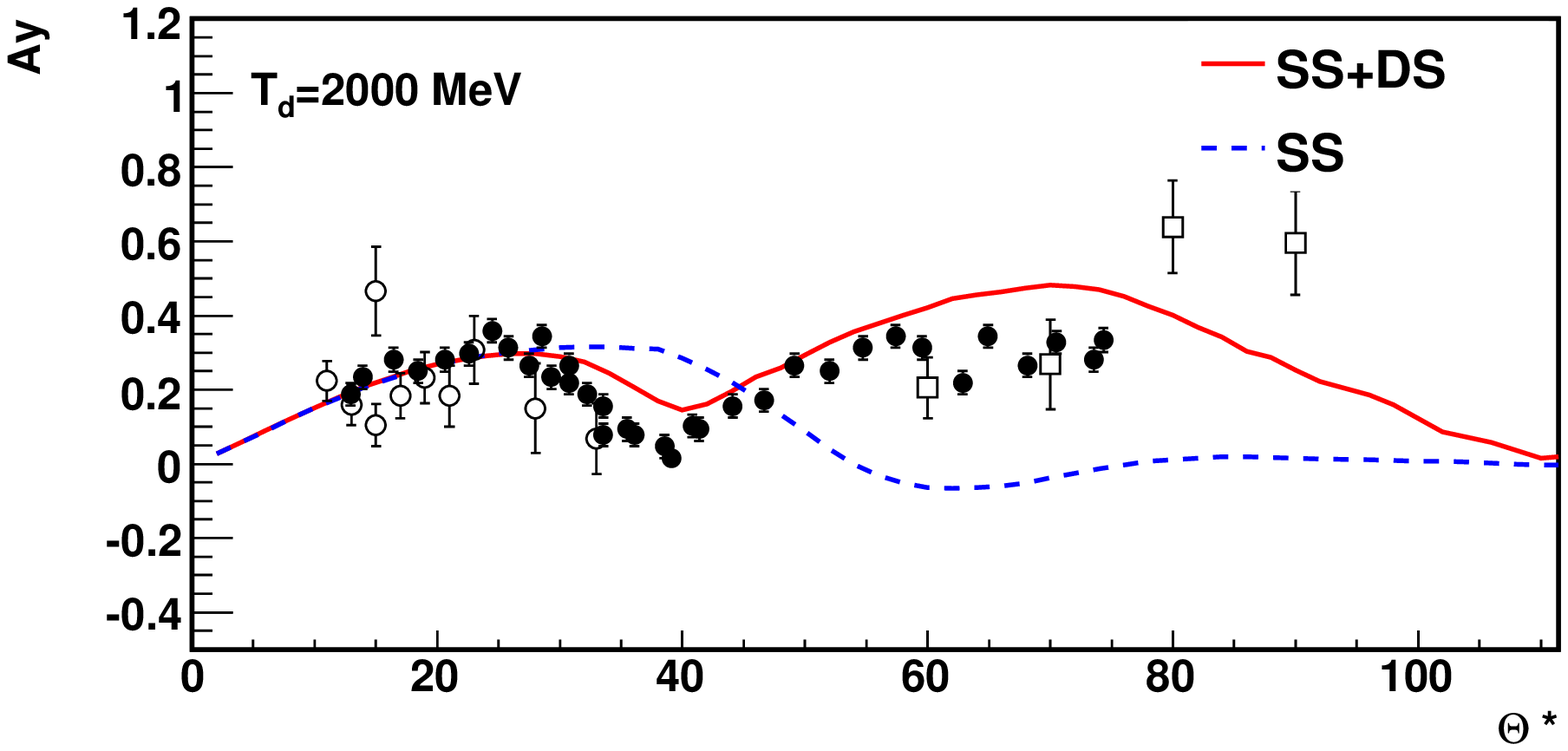}}
    \caption{Vector $A_y$ analyzing power in $dp$-elastic scattering at 2000 MeV. The symbols and curves are explained in the text.}
    \label{fig:ay_2000MeV}
\end{minipage}
\begin{minipage}[htbp]{0.48\textwidth}
\vspace*{-1.18cm}
%\begin{figure}[htb]
  \centering
    % please do not add file name extension this makes switching between latex and pdflatex easier
    \resizebox{0.97\textwidth}{!}{\includegraphics{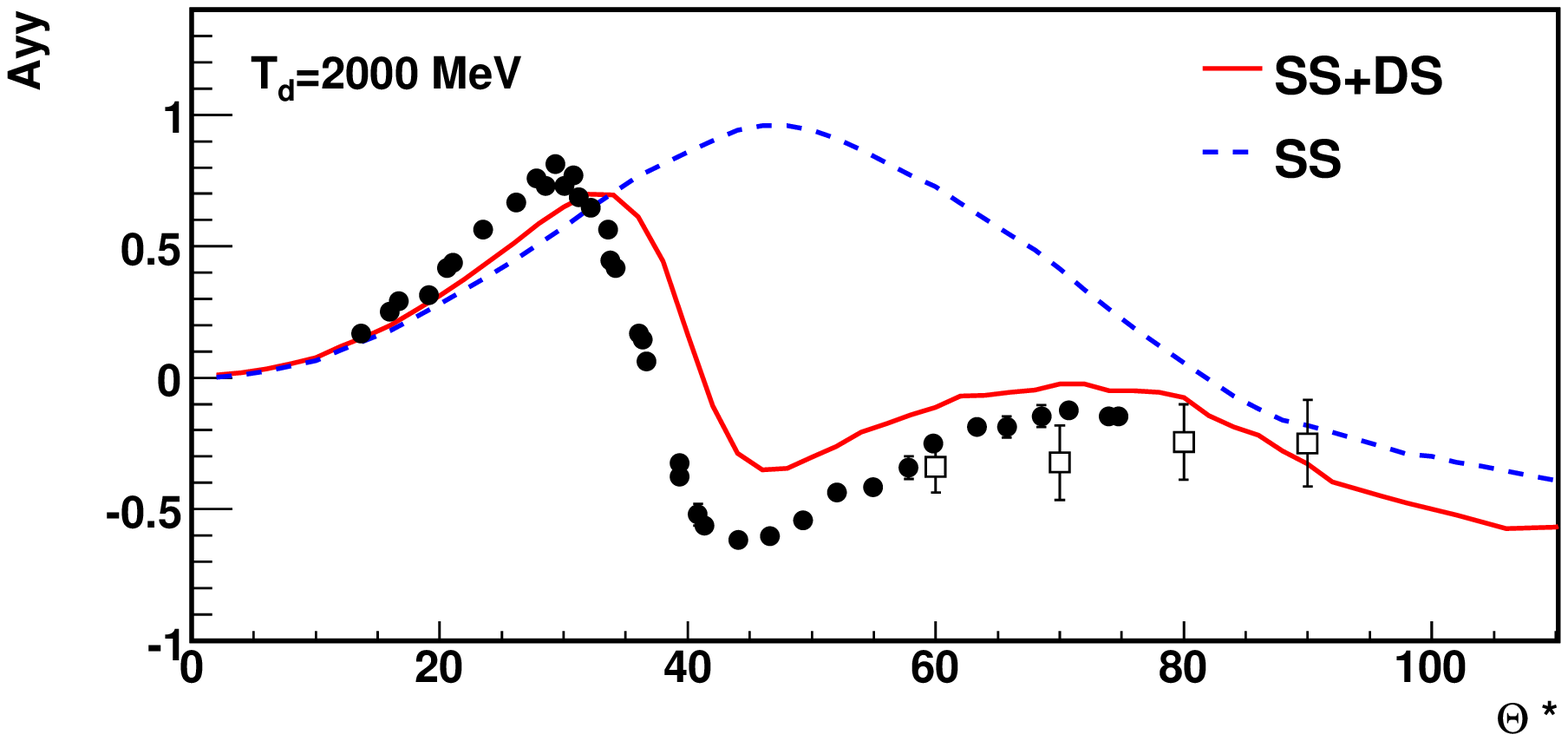}}
    \caption{Tensor $A_{yy}$ analyzing power in $dp$-elastic scattering at 2000 MeV. The symbols and curves are the same as in Figure~\ref{fig:ay_2000MeV}.}
    \label{fig:ayy_2000MeV}
\end{minipage}
\end{figure}
The results on the angular dependence of the vector $A_{y}$ and tensor $A_{yy}$ analyzing powers in $dp-$ elastic scattering at 2000 MeV are shown in Figures~\ref{fig:ay_2000MeV} and ~\ref{fig:ayy_2000MeV}, respectively. The data obtained at Argonne National Laboratory(ANL) are presented by the solid symbols. Open squares and circles are the data obtained at the ITS and at hydrogen bubble chamber at JINR, respectively. The dashed and solid lines are the results of the relativistic multiple scattering model calculations\cite{nb_ladygina} with and without of the double scattering term. The full calculations are in a  reasonable agreement with the data.\\
The dependencies of the tensor $A_{yy}$ analyzing power in $dp-$ elastic scattering obtained at the fixed angles of 60$^\circ$, 70$^\circ$, 80$^\circ$ and 90$^\circ$ in the cms as a function of transverse momentum $p_{T}$ are shown in Figure~\ref{fig:ayy_versus_pt}. The open and solid symbols represent the data obtained at RIKEN, Saclay, ANL and at the Nuclotron, respectively. It would be interesting to extend the range of the measurements to larger $p_T$, where the manifestation of non-nucleonic degrees of freedom is expected.\\
%%%%%%%%%%%%%%%%%%%%%%%%%%%%%%%%%%%%%%%%%%%%%%%%%%%%%
\begin{figure}[t]
\begin{minipage}[t]{0.48\textwidth}
  \centering
    % please do not add file name extension this makes switching between latex and pdflatex easier
    \resizebox{0.90\textwidth}{!}{\includegraphics[width=0.8\textwidth,height=0.59\textwidth]{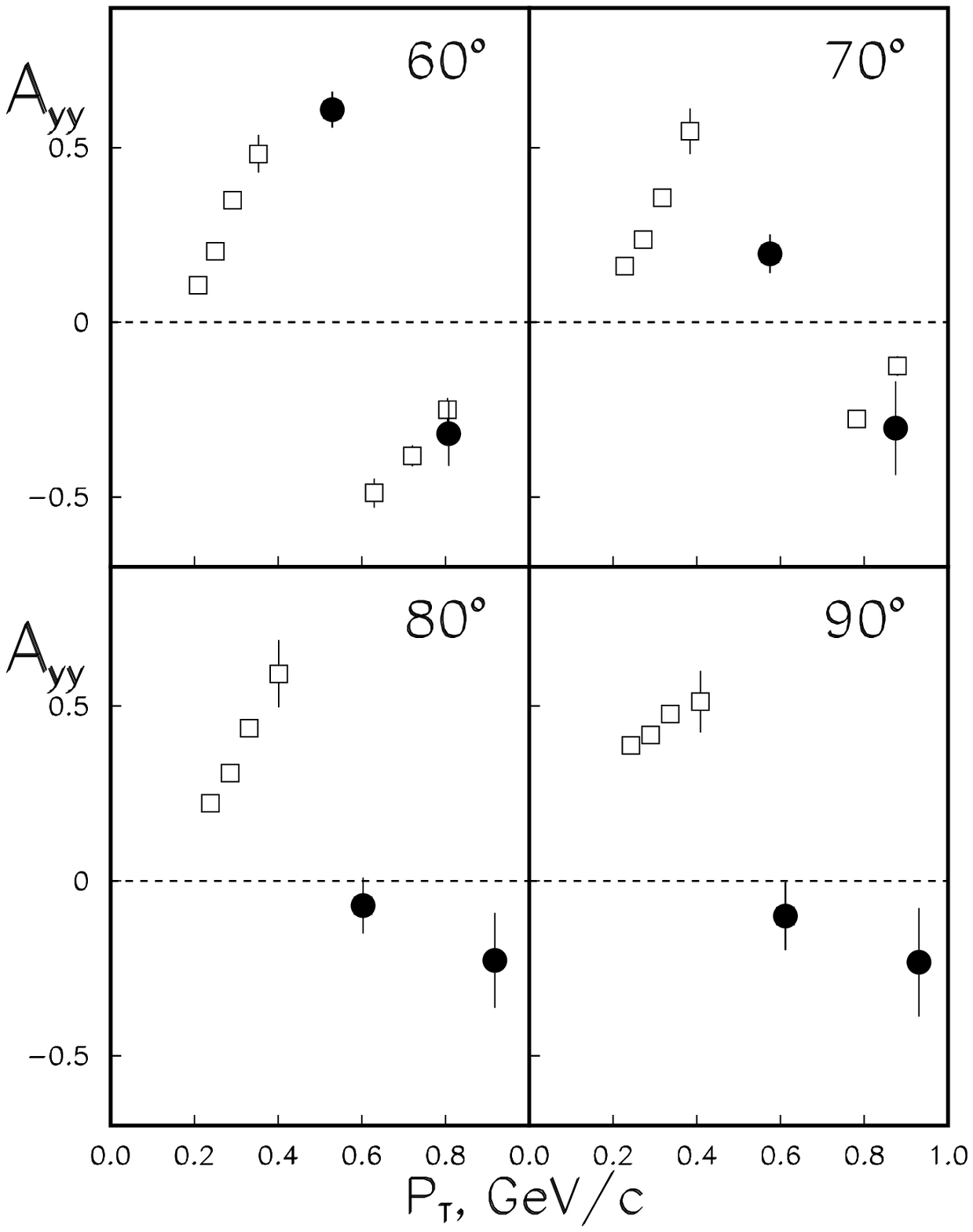}}
    \caption{Tensor $A_{yy}$ analyzing power in $dp$-elastic scattering obtained at the fixed angles of 60$^\circ$, 70$^\circ$, 80$^\circ$ and 90$^\circ$ in cms as a function of transverse momentum $p_T$. The symbols are explained in text.}
    \label{fig:ayy_versus_pt}
\end{minipage}
\begin{minipage}[htbp]{0.48\textwidth}
\vspace*{-2.1cm}
%\begin{figure}[htb]
  \centering
    % please do not add file name extension this makes switching between latex and pdflatex easier
    \resizebox{0.9\textwidth}{!}{\includegraphics[width=0.55\textwidth,height=0.41\textwidth]{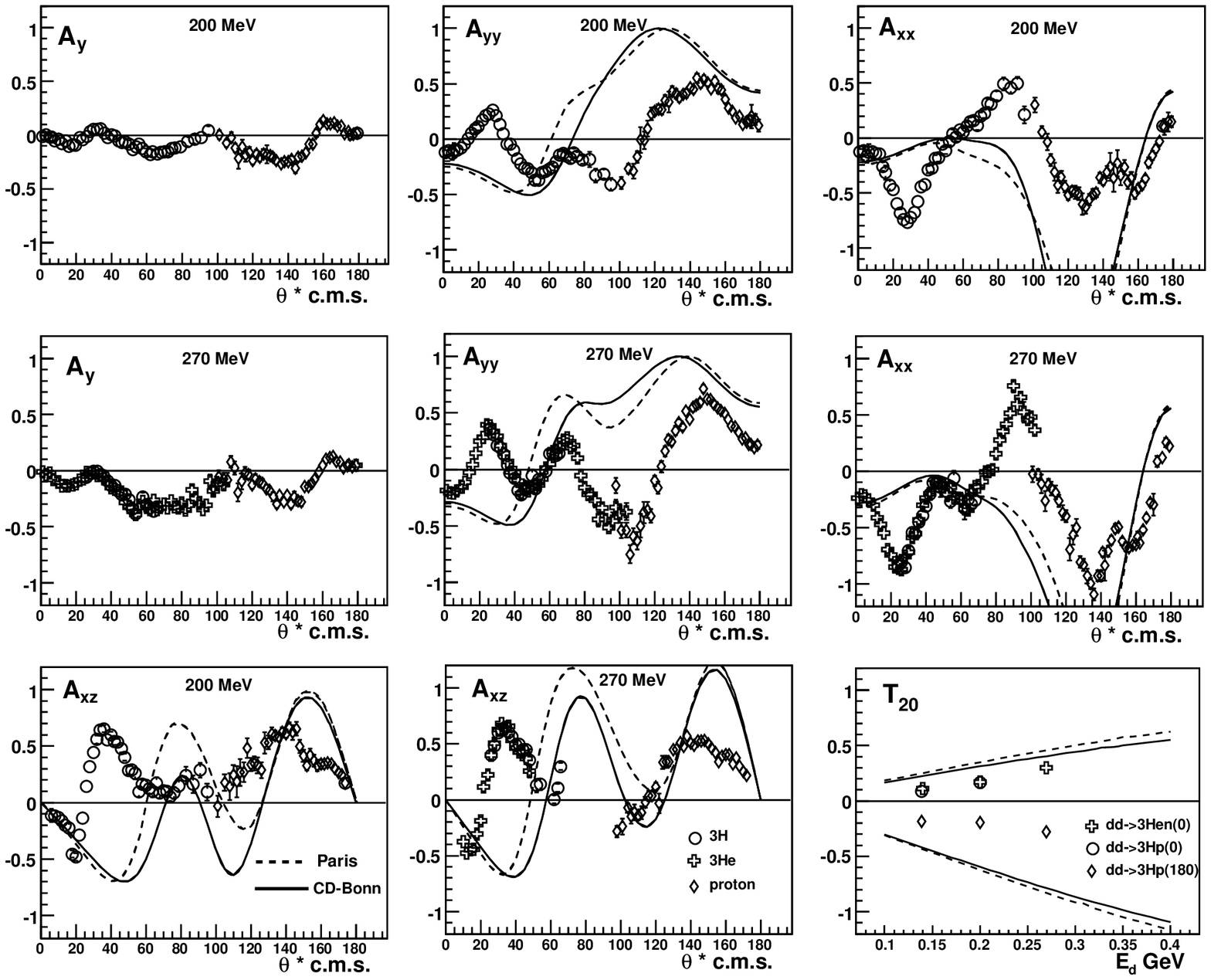}}
    \caption{The analyzing powers data in d(d,$^3He$)n and d(d,$^3H$)p at 200 and 270 MeV. The curves are the calculation within one-nucleon-exchange approximation.}
    \label{fig:riken_data}
\end{minipage}
\end{figure}
Figure~\ref{fig:riken_data} presents the analyzing powers results in the d(d,$^3$H)p and d(d,$^3$He)n reactions obtained at RARF(RIKEN, Japan) at 200 and 270 MeV. The details of the experiment can be found in \cite{RIKEN_experiment}. The solid and long-dashed curves are the result of ONE calculations using CD-Bonn and Paris deuteron and $^3$He wave function, respectively. One can see that ONE calculations are in the qualitative agreement with the data on the $T_{20}$. %The behavior of the tensor analyzing powers $A_{yy}$, $A_{xx}$ and $A_{xz}$ are not reproduced by the ONE model. The ONE model predicts a zero value of the vector analyzing powers $A_y$. Some structures in the experimental results on $A_y$ indicate on the possibility other than ONE mechanisms in these reactions. The reason of these discrepancy can be in the non-adequateness of the 3N-bound state spin structure and/or more complicated reaction mechanism. The multiple scattering calculations are in progress now.
The analyzing powers behaviors are not reproduced by ONE model. The reason of this discrepancy can be in the inadequate description of the $3N$-bound state spin structure and/or more complicated reaction mechanism. The multiple scattering calculations are in progress now.  
\section{Future plans }
Future plans of DSS (Deuteron spin structure) - collaborations in spin studies are related with the construction of new polarized deuteron source. %This source will provide the intensity up to 2$\times$10$^{10}$ ppp and larger variety of the spin modes than POLARIS. Figure of merit of new source will be increased by a factor $10^3$ compared with POLARIS. 
The energy scan of the $dp-$elastic scattering observables and measurements of the analyzing powers in $dp-$ nonmesonic breakup will be done using internal target and polarized deuteron beam from new PIS. 
%%%%%%%%%%%%%%%%%%%%%%%%%%%%%%%%%%%%%%%%%%%%%%%%%%%%%
%%%%%%%%%%%%%%%%%%%%%%%%%%%%%%%%%%%%%%%%%%%%%%%%%%%%%%%%%%%%%%%%%%%%%%%%%%%%%
%%%%%%%%%%%%%%%%%%%%%%%%%%%%%%%%%%%%%%%%%%%%%%%%%%%%%%%%%%%%%%%%%%%%%%%%%%%%%%%%%
%%%%%%%%%%%%%%%%%%%%%%%%%%%%%%%%%%%%%%%%%%%%%%%%%%%%%%%%%%%%%%%%%%%%%%%%%%%%%%%%%
% the recommended way to include figures
\begin{figure}[t]
\begin{minipage}[t]{0.45\textwidth}
  \centering
    % please do not add file name extension this makes switching between latex and pdflatex easier
    \resizebox{0.8\textwidth}{!}{\includegraphics[width=0.8\textwidth,height=0.55\textwidth]{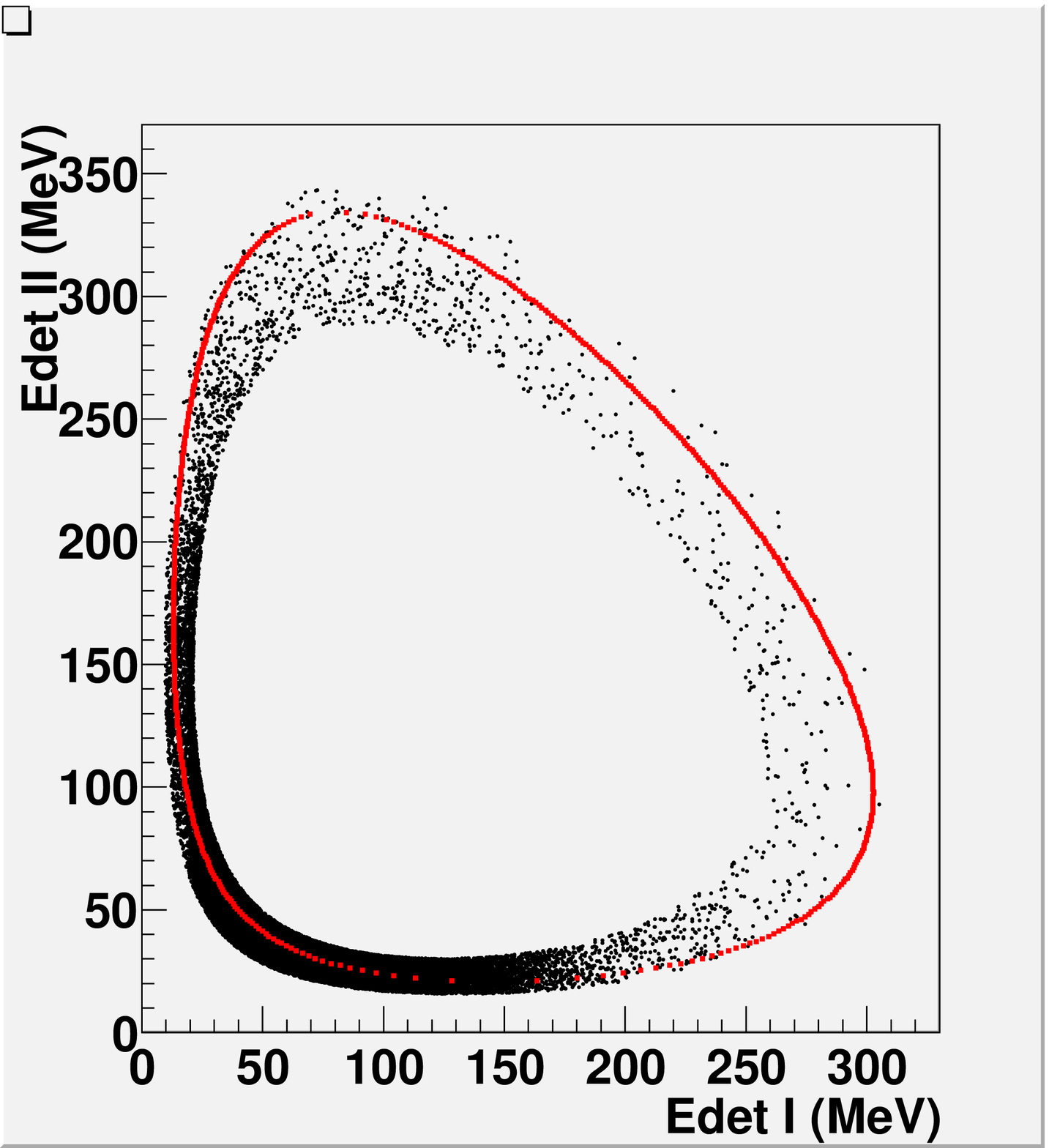}}
    \caption{The correlation of the $\Delta E+E$ information from 2 proton detectors in case the dp-breakup reaction investigation at 500 MeV. $\Theta_{1}=34^\circ$, $\Theta_{2}=29.8^\circ$, $\phi_{12}=180^\circ$}
    \label{fig:dp_breakup}
\end{minipage}
\begin{minipage}[htbp]{0.45\textwidth}
\vspace*{-0.50cm}
%\begin{figure}[htb]
  \centering
    % please do not add file name extension this makes switching between latex and pdflatex easier
    \resizebox{0.88\textwidth}{!}{\includegraphics[width=0.8\textwidth,height=0.55\textwidth]{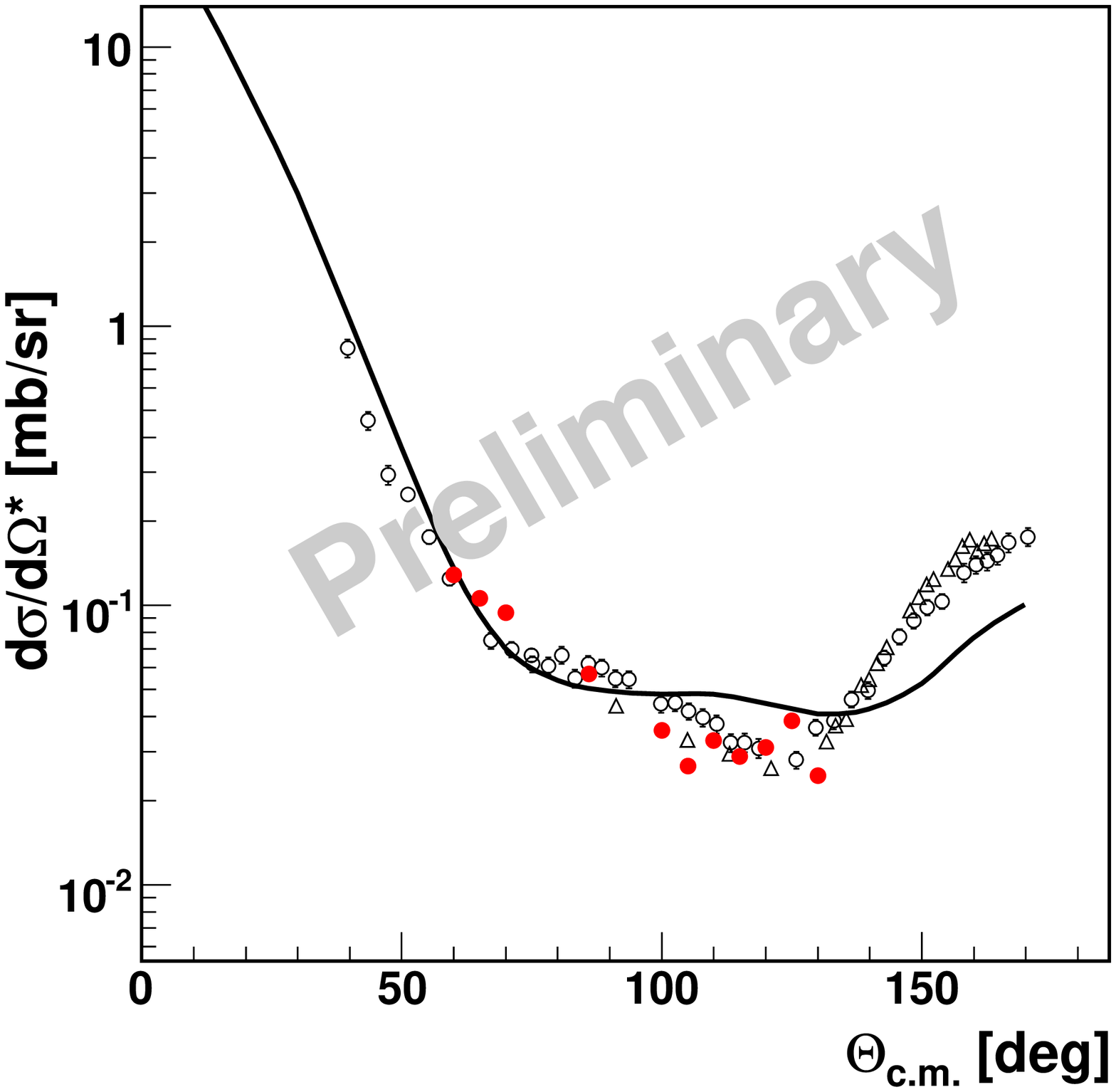}}
    \caption{The results on the angular dependence of the dp-elastic cross section obtained at 880 MeV at Nuclotron in March 2011. World data at 850 MeV and 940 MeV are marked by the open triangles and circles, respectively.}
    \label{fig:dp_cs_880}
\end{minipage}
\end{figure}
%
%%%%%%%%%%%%%%%%%%%%%%%%%%%%%%%%%%%%%%%%%%%%%%%%%%%%%%%%%%%%%%%%%%%%%%%%%%%%%%%%%
The $dp-$elastic scattering and $dp-$nonmesonic breakup cross section measurements can be done with the current unpolarized ion source as the first step. The $dp-$nonmesonic breakup reaction will be investigated at ITS at the Nuclotron using $\Delta E-E$ techniques for the detection of two final protons. Figure~\ref{fig:dp_breakup} presents the correlation of the $\Delta E-E$ information from 2 proton detectors. A kinematic relation are shown by the solid line. The preliminary results on the angular dependence of the $dp-$elastic scattering cross section obtained at 880 MeV at the Nuclotron in March 2011 are presented in Figure~\ref{fig:dp_cs_880} by the solid symbols. They are compared with experimental data obtained at 850 MeV and 940 MeV given by the open triangles and circles, respectively. Solid line are the result multiple scattering model calculations\cite{nb_ladygina}.\\% Normalization of the Nuclotron data are relatively only.\\ 
The first line experiment with the extracted polarized deuteron beam for new PIS is the spin observables study for the $^3H(d,p)^4He$ reaction at the energies 1.0-1.75 GeV, where the contribution from the deuteron D-state is expected to reach the maximum.\\
New experimental data will ensure the important information about the light nuclei spin structure at short internucleonic distances, where the relativistic effects and 3N forces play an important role.\\

\acknowledgements{%
\vspace*{-0.3cm}
  The work was supported in part by the RFBR under Grant $N^\circ 10-02-00087a$.
} 
%%%%%%%%%%%%%%%%%%%%%%%%%%%%%%%%%%%%%%%%%%%%%%%%%%%%%%%%%%%%%%%%%%%%%%%%%%%%%%%%%
%%%%%%%%%%%%%%%%%%%%%%%%%%%%%%%%%%%%%%%%%%%%%%%%%%%%%%%%%%%%%%%%%%%%%%%%%%%%%%%%%
% bibliographic items can be constructed using the LaTeX format in SPIRES
% see http://www.slac.stanford.edu/spires/hep/latex.html
% SPIRES will also supply the CITATION line information; please include it

%
%%%%%%%%%%%%%%%%%%%%%%%%%%%%%%%%%%%%%%%%%%%%%%%%%%%%%%%%%%%%%%%%%%%%%%%%%%%%%%%%%

}  % do not remove

%%% Local Variables: 
%%% mode: latex
%%% TeX-master: "../hadron2011.tex"
%%% End: 